\documentclass[12pt]{article}
\usepackage{epsfig}
\usepackage{latexsym}

\newcommand\pubnumber{SLAC--PUB--10747 }
\newcommand\pubdate{\today}
\newcommand\hepnumber{hep-th/0409192}

\def\SLAC{Stanford Linear Accelerator Center\\
    Stanford University, Stanford, California 94025 USA}
\def\doeack{\footnote{Work supported by the Department of Energy,
                     contract DE--AC02--76SF00515.}}

\def\Title#1{\begin{center} {\Large #1 } \end{center}}
\def\Author#1{\begin{center}{ \sc #1} \end{center}}
\def\Address#1{\begin{center}{ \it #1} \end{center}}

\newcommand\pubblock{\rightline{\begin{tabular}{l} \pubnumber\\
         \pubdate \\ \hepnumber \end{tabular}}}
\newenvironment{Abstract}{\begin{quotation} \begin{center}
                       ABSTRACT
     \end{center}\bigskip  }{\end{quotation}}

\def\Acknowledgements{\bigskip  \bigskip \begin{center} \begin{large}
             \bf Acknowledgements \end{large}\end{center}}
\def\beq{\begin{equation}}
\def\eeq#1{\label{#1}\end{equation}}
\def\eeqn{\end{equation}}

\newenvironment{Eqnarray}%
   {\arraycolsep 0.14em\begin{eqnarray}}{\end{eqnarray}}
\def\beqa{\begin{Eqnarray}}
\def\eeqa#1{\label{#1}\end{Eqnarray}}
\def\eeqan{\end{Eqnarray}}

\let\bar=\overbar

\def\lsim{\mathrel{\raise.3ex\hbox{$<$\kern-.75em\lower1ex\hbox{$\sim$}}}}
\def\gsim{\mathrel{\raise.3ex\hbox{$>$\kern-.75em\lower1ex\hbox{$\sim$}}}}

\def\del{\partial}
\def\Dslash{\not{\hbox{\kern-4pt $D$}}}
\def\dslash{\not{\hbox{\kern-2pt $\del$}}}

\def\msb{{\bar{\scriptsize M \kern -1pt S}}}

\def\drb{{\bar{\scriptsize D \kern -1pt R}}}

\makeatletter
\def\section{\@startsection{section}{0}{\z@}{5.5ex plus .5ex minus
 1.5ex}{2.3ex plus .2ex}{\large\bf}}
\def\subsection{\@startsection{subsection}{1}{\z@}{3.5ex plus .5ex minus
 1.5ex}{1.3ex plus .2ex}{\normalsize\bf}}
\def\subsubsection{\@startsection{subsubsection}{2}{\z@}{-3.5ex plus
-1ex minus  -.2ex}{2.3ex plus .2ex}{\normalsize\sl}}

\renewcommand{\@makecaption}[2]{%
   \vskip 10pt
   \setbox\@tempboxa\hbox{\small #1: #2}
   \ifdim \wd\@tempboxa >\hsize     % IF longer than one line:
       \small #1: #2\par          %   THEN set as ordinary paragraph.
     \else                        %   ELSE  center.
       \hbox to\hsize{\hfil\box\@tempboxa\hfil}
   \fi}

 \def\citenum#1{{\def\@cite##1##2{##1}\cite{#1}}}
\newcount\@tempcntc
\def\@citex[#1]#2{\if@filesw\immediate\write\@auxout{\string\citation{#2}}\fi
  \@tempcnta\z@\@tempcntb\m@ne\def\@citea{}\@cite{\@for\@citeb:=#2\do
    {\@ifundefined
       {b@\@citeb}{\@citeo\@tempcntb\m@ne\@citea\def\@citea{,}{\bf ?}\@warning
       {Citation `\@citeb' on page \thepage \space undefined}}%
    {\setbox\z@\hbox{\global\@tempcntc0\csname b@\@citeb\endcsname\relax}%
     \ifnum\@tempcntc=\z@ \@citeo\@tempcntb\m@ne
       \@citea\def\@citea{,}\hbox{\csname b@\@citeb\endcsname}%
     \else
      \advance\@tempcntb\@ne
      \ifnum\@tempcntb=\@tempcntc
      \else\advance\@tempcntb\m@ne\@citeo
      \@tempcnta\@tempcntc\@tempcntb\@tempcntc\fi\fi}}\@citeo}{#1}}
\def\@citeo{\ifnum\@tempcnta>\@tempcntb\else\@citea\def\@citea{,}%
  \ifnum\@tempcnta=\@tempcntb\the\@tempcnta\else
  {\advance\@tempcnta\@ne\ifnum\@tempcnta=\@tempcntb \else\def\@citea{--}\fi
    \advance\@tempcnta\m@ne\the\@tempcnta\@citea\the\@tempcntb}\fi\fi}
\makeatother

\begin{document}
\begin{titlepage}
\pubblock

\vfill
\Title{The Wilsonian Renormalization Group in Randall-Sundrum 1}
\vfill
\Author{Adam Lewandowski \doeack}
\Address{\SLAC}
%\andauth
%\Address{\napoli}
\vfill
\begin{Abstract}
We find renormalization group transformations for the compactified Randall-Sundrum
scenario by integrating out an infinitesimal slice of ultraviolet degrees of freedom near
the Planck brane.
Under these transformations the coefficients of operators on the Planck brane experience 
RG evolution.  The extra-dimensional radius also scales, flowing to zero in the IR.  
We find an attractive fixed point in the context of a bulk scalar field theory.  
Calculations are simplified in the low energy
 effective theory as we demonstrate with the computation of a loop diagram.
\end{Abstract}
\vfill
\end{titlepage}
\def\thefootnote{\fnsymbol{footnote}}
\setcounter{footnote}{0}

\section{Introduction}

Effective field theories in anti-de Sitter space have received much attention with their
application to the weak/Planck hierarchy problem in the compactified Randall-Sundrum 
scenario (RS1) \cite{rs1}.  However, because
of the warped geometry, loop calculations are challenging in this
picture.  It is difficult even to power count the size of 
diagrams because of the wide range of scales appearing in loops.  
In contexts where these effects are important such as
the stability of the hierarchy and gauge unification~\cite{adshard} one is forced
to confront these difficulties. 

In Ref. \cite{runrs}
holographic RG ideas \cite{holorg,skenderis} were developed
to obtain a low energy effective field theory description of RS1 with a smaller 
extra-dimensional radius.  Further developments are found in
\cite{runrsothers,higherspin}.  By integrating out short distance {\it classical} degrees of 
freedom couplings were found to experience RG evolution with the size of
the extra-dimension.  The large
warp factor appears in ways predicted by the running of couplings near a fixed point
as they flow with the size of the extra dimension simplifying power counting.  
 In this paper, we develop this approach further integrating out short distance classical and 
 {\it quantum} degrees of freedom to find renormalization group transformations.  This has
  the features of a Wilsonian integration as applied to the RS1 scenario.
 
We will consider an orbifold even scalar field in the 5D warped two brane geometry of 
RS1.  The metric is
\begin{equation}~\label{metric}
ds^2 = \frac{1}{(kz)^2} ( \eta_{\mu \nu} dx^{\mu} dx^{\nu} - dz^2)
\end{equation}
where $z$ parameterizes the extra-dimensional space  and $1/k$ is the radius of curvature.  
The space is bounded by a UV (Planck) brane at $z=z_{UV} = 1/k = 1/{\cal O}(M_{Pl})$
and an IR  (TeV) brane at $z=z_{IR} = 1/{\cal O}({\rm TeV})$.  We will work in a mixed
$z$-position/$x$-momentum space basis for the scalar and implement a physical hard 
4-momentum cutoff.  

Because of the warping, the shortest distance degrees of freedom are located in 
the region near the UV brane.   To implement the RG we will integrate out a sliver 
of these 
short distance modes.  The result of the integration is to modify the couplings of
operators on the UV brane.  We will make this transformation continuous, finding
$\beta$ functions for UV brane couplings.  Upon RG evolution to the IR the 
interbrane distance flows to zero.   Correlation 
functions of the scalar field in the low scale effective theory match to
correlation functions evaluated in the high scale theory in the region of overlap of the two 
theories. Renormalization group flows in AdS space were also studied
in \cite{systematics} and by dimensional deconstruction \cite{deconstruction}
in \cite{deconstructads}.
Although we do not employ the AdS/CFT correspondence \cite{adscft} certain features of 
the CFT dual of RS \cite{rscft} are exhibited.

The developments of this paper allow for the calculation of loop diagrams in RS1 in a 
low energy effective theory with a smaller extra-dimensional radius.  
In particular, in sufficiently low energy theories where
 this radius is effectively zero there is no warping
and diagrams are straightforward to calculate.    At such low energies the propagator
simplifies and the cutoff does not vary significantly.   Because of the presence of
an attractive fixed point, RG evolution is easy for perturbations near this fixed point.
In this paper, we give an example of a loop calculation in the low energy effective field
theory where we make use of these simplifications.  

The paper is organized as follows.  In Section 2 we outline the RG as applied
to scalar field theory in the RS1 background.  In Section 3 we derive $\beta$
 functions for couplings of operators on the UV brane.  
 The fixed point behavior is discussed in Section 4.  
In Section 5 we present an example calculation
of a loop in the low energy effective theory.  We discuss and summarize our 
results in Section 6.

\section{Overview of the RG in RS1}

In this section we will outline the procedure to find RG flows 
for fields in a fixed RS1 background.
   
We consider theories with actions of the form 
\begin{equation}
S= S_{bulk} + S_{UV} +  S_{IR},
\end{equation}
where
\begin{eqnarray}~\label{action1}
S_{bulk}[\chi_i(x,z)] & =  &\int d^4 x \int_{z_{UV}}^{z_{IR}} dz 
\sqrt{G} \left( \sum_i C^i_{bulk} {\cal O}^i_{bulk} \right), \nonumber \\
S_{UV}[\chi_i(x,z_{UV})] & = & \int d^4 x \int_{z_{UV}}^{z_{IR}} dz \sqrt{g}
 \left( \sum_i C^i_{UV} {\cal O}^i_{UV} \right) \delta(z-z_{UV}), \nonumber \\
S_{IR}[\chi_i(x,z_{IR})] & = & \int d^4 x \int_{z_{UV}}^{z_{IR}} dz \sqrt{g} 
\left( \sum_i C^i_{IR} {\cal O}^i_{IR} \right)  \delta(z-z_{IR}).
\end{eqnarray}
where $G^{MN}$ is the AdS bulk metric defined in (\ref{metric}) 
and $g^{\mu \nu}(x) \equiv G^{\mu \nu}(x, z)$. 
 The local operators ${\cal O}_{UV}$ and ${\cal O}_{IR}$ contain products
 of the scalar fields, $\chi_i(x,z)$, and their $x^{\mu}$ derivatives contracted with $g^{\mu \nu}$.  
The bulk operators ${\cal O}_{bulk}$ are local operators of $\chi_i$ and derivatives with respect to
$\partial_M$ contracted with $G^{MN}$.  Here and throughout this
paper we work in units of the curvature $k$ setting $k=1$.

To understand the Wilsonian RG (for a review see Ref. \cite{wilson}) 
as applied here imagine first a
 5D lattice version of this theory.  The warped geometry dictates that the 
lattice spacing is most dense in both $x$ and $z$ directions in the region near the UV
brane becoming exponentially less dense as we move towards the IR brane.  In the real
space Wilsonian picture we wish to integrate out a subset of the sites to arrive
at a self-similar action with larger lattice spacing.  In RS1 the sites spaced
by the fundamental lattice spacing are found only in the UV brane region.  If we
thin the lattice by integrating out a subset of these sites the result is to move the 
UV boundary towards the IR and modify the lattice couplings at sites at the UV boundary.  
There will be no modification of couplings at sites away from this boundary.

We will not employ a real space picture but the features of this RG will be present.  We
will work in the continuum in 
a mixed 4-momentum $p$, $z$-position space basis.  This will allow us to find
smooth RG transformations.  We write the fields $\chi_i(x,z)$ in 
4-momentum components, $\chi_i(p,z)$ and integrate out a sliver of the modes with
highest 
4-momentum and a sliver of the $z$-position modes local to the UV brane.  

We will employ a physical 4-momentum cutoff.
In our coordinate system the scale symmetry of AdS takes the form $x^{\mu} \to \lambda x^{\mu}$,
($p^{\mu} \to p^{\mu} / \lambda$) and 
$z \to \lambda z $.  To resepct this symmetry it is necessary that the cutoff depend on 
position as $\Lambda/z$.  The cutoff is maximal at the location of the UV brane where it is 
$\Lambda / z_{UV}$.  (Though $z_{UV}$=1
in our $k=1$ units we will keep factors explicit where needed for clarity).

We first integrate out those highest momentum modes in a sliver near the maximal
cutoff $(1-\epsilon) \Lambda /z_{UV} < |p| \le \Lambda /z_{UV}$ with $\epsilon$ a
 small positive parameter.  These
modes are in the $z$-space region $z_{UV} \le z < z_{UV}(1+\epsilon)$.  
If the momentum sliver is small enough
($\epsilon$ is sufficiently close to 0) we can
write the result of this integration as a modification to the coefficients of operators on the 
UV brane.  To first order in $\epsilon$ we may write this transformation as
 \begin{equation}
 S_{UV} \to S'_{UV} = \int d^4 x \int_{z_{UV}}^{z_{IR}} dz \sqrt{g}
 \left( \sum_i (C^i_{UV} + \Delta C^i_{mom}) {\cal O}^i_{UV} \right) 
 \delta(z-z_{UV}).
\end{equation}
After the integration the cutoff is a flat $(1-\epsilon) \Lambda/z_{UV}$ in the region 
$z_{UV} \le z < z_{UV}(1+\epsilon)$.

Consistent with the scale symmetry of the space and self-similarity of the effective theory
we must  integrate out modes of $z$
space position in the region $z_{UV} \le z < z_{UV}(1+\epsilon)$.  
Upon performing the integration the theory will have a new boundary of space at 
$z=z_{UV}(1+\epsilon)$ with an effective UV brane action at this boundary.  
The UV brane will now have the form
\begin{equation}
S^{\prime \prime}_{UV} = \int d^4 x \int_{z_{UV}}^{z_{IR}} dz \sqrt{g}
 \left( \sum_i (C^i_{UV} + \Delta C^i_{mom} + \Delta C^i_{z-pos}) {\cal O}^i_{UV} 
 \right) \delta(z-z_{UV} (1+\epsilon)),
\end{equation}

To complete the RG transformation we rescale coordinates $x \to (1-\epsilon) x$, 
($p \to p(1+\epsilon)$) and
 $z \to (1-\epsilon) z$.  This will bring our integrated theory into a from
 self-similar to (\ref{action1}).   Both bulk and brane operators
 are invariant under this rescaling because of the scale symmetry of AdS.  
 They transform as ${\cal O} \to {\cal O}$.   The effective theory then takes the form
\begin{eqnarray}
S_{bulk}^{eff}[\chi_i(x,z)] & =  &\int d^4 x \int_{z_{UV}}^{z'_{IR}} dz 
\sqrt{G} \left( \sum_i C^{'i}_{bulk} {\cal O}^i_{bulk} \right), \nonumber \\
S_{UV}^{eff}[\chi_i(x,z_{UV})] & = & \int d^4 x \int_{z_{UV}}^{z'_{IR}} dz \sqrt{g}
 \left( \sum_i C^{'i}_{UV} {\cal O}^i_{UV} \right) \delta(z-z_{UV}), \nonumber \\
S_{IR}^{eff}[\chi_i(x,z_{IR})] & = & \int d^4 x \int_{z_{UV}}^{z'_{IR}} dz \sqrt{g} 
\left( \sum_i C^{'i}_{IR} {\cal O}^i_{IR} \right)  \delta(z-z'_{IR}).
\end{eqnarray}
Here the momentum cutoff is  $\Lambda/z$  and the UV brane
boundary is restored to $z_{UV}$.  The primed
parameters are
\begin{eqnarray}
C^{'i}_{bulk} & = & C^{i}_{bulk}, \nonumber \\
C^{'i}_{UV} & = & C^i_{UV} + \Delta C^i_{mom} + \Delta C^i_{z-pos}, \nonumber \\
C^{'i}_{IR} & = & C^{i}_{IR}, \nonumber \\
z'_{IR} & = &  z_{IR} - \epsilon z_{IR}.
\end{eqnarray}

We see that both the bulk and IR brane coefficients do not transform because
of the scale invariance of the theory.  The
UV brane coefficients flow in a way determined by the physics integrated out near the
UV brane.  The scaling behaviour depends significantly on the form these corrections
take.  The inverse of the IR brane position, $1/z_{IR}$, flows as a relevant mass parameter.
Because of this the effect of successive iterations of this integration procedure
 is to decrease the interbrane distance.   We will examine the continuous form of this RG
 in the next section by taking $\epsilon \to 0$.

\section{RG Flows in Detail}

In this section we obtain 
$\beta$ functions for UV brane coefficients focusing
 on a single orbifold even scalar, $\chi$, in an RS1 background.
We work with an action having the form
\begin{equation}
S= S_{bulk} + \frac{1}{2} S_{UV} + \frac{1}{2} S_{IR},
\end{equation}
where
\begin{eqnarray}~\label{action2}
S_{bulk}[\chi(x,z)] & =  &\int d^4 x \int_{z_{UV}}^{z_{IR}} dz 
\sqrt{G} \left( \frac{1}{2} G^{MN} 
\partial_M \chi \partial_N \chi - V_{bulk} (\chi) \right), \nonumber \\
S_{UV}[\chi(x,z_{UV})] & = & \int d^4 x \sqrt{g_{UV}} L_{UV}(\chi(x, z_{UV}), 
\partial_{\mu}, 
g_{UV}^{\mu \nu}), \nonumber \\
S_{IR}[\chi(x,z_{IR})] & = & \int d^4 x \sqrt{g_{IR}} L_{IR} (\chi(x, z_{IR}), \partial_{\mu}, 
g_{IR}^{\mu \nu}).
\end{eqnarray}
where $g_{IR,UV}^{\mu \nu}(x) \equiv G^{\mu \nu}(x, z=z_{IR,UV})$.  $L_{UV}$ and 
$L_{IR}$ are local
functions of $\chi$ and its $x^{\mu}$ derivatives.  We implement a position dependent 
physical 4-momentum cutoff $\Lambda /z$.

%Correlation functions of $\chi$ are given by
%\begin{equation}
%\langle \chi(x_1,z_1) \ldots \chi(x_n,z_n) \rangle = 
%\frac{ \int [d\chi] \chi(x_1,z_1) \ldots \chi(x_n,z_n) e^{i S[\chi]}}{\int [d \chi]
%e^{i S[\chi]}}.
%\end{equation}
%The integration measure is
%\begin{equation}
%[d \chi ] = \prod_{z_i} \prod_{|k_j| <\Lambda / z_i } d \chi (k_j,z_i),
%\end{equation}
%which implements a position dependent hard momentum cutoff at $\Lambda /z$.

We will integrate out an infinitesimal sliver of scalar modes
in the region $z_{UV} \le z < z_{UV}(1+\epsilon)$. We perform a path
integral over these modes in two steps.  We first integrate out components of the field 
with 4-momentum, $p$, in
the range $ \Lambda /z_{UV}(1+\epsilon) < |p| \le \Lambda /z_{UV}$ 
then integrate out the remaining components with $z$ values 
$z_{UV} \le z < z_{UV}(1+\epsilon )$.

To integrate out the high momentum modes we write the field in components
\begin{equation}
\chi(x,z) = \chi_{<}(x,z) + \chi_{>}(x,z), 
\end{equation}
where $\chi_{<}$ contains momentum modes $|p| \le \Lambda / z_{UV}(1+\epsilon)$ and 
$\chi_{>}$ contains modes with momentum $\Lambda / z_{UV}(1+\epsilon) < |p| \le 
\Lambda / z_{UV}$.  Because these modes are confined to an epsilonic region
near the effective UV
brane, to ${\cal O}(\epsilon)$ we can neglect the contribution
to the bulk action and write
\begin{equation}
S[\chi] = S_{IR}[\chi_{<}] + S_{bulk}[\chi_{<}] + S_{UV}[\chi_{<} + \chi_{>}].
\end{equation}
We let
\begin{equation}
e^{i (S_{IR}[\chi_<]+S_{bulk}[\chi_<]+S'_{UV}[\chi_<])} = 
\int [d \chi_>] e^{i S[\chi]},
\end{equation}
and perform the path integral. 

We may integrate out $\chi_{>}$ in perturbation theory recognizing that
\begin{equation}
\langle \chi_{>}(x,z_{UV}) \chi_{>}(x',z_{UV}) \rangle = 
\int \frac{d^4 p}{(2 \pi)^4} 
(-i) \Delta_p(z_{UV},z_{UV}) \Theta(p) e^{i p \cdot (x-x')},
\end{equation}
where
\begin{equation}
\Theta(p) = \left\{ 
\begin{array}{ll} %
1 & \Lambda / z_{UV}(1+\epsilon) < |p| < \Lambda / z_{UV} \\
0 &  {\rm  otherwise}
\end{array} \right.
\end{equation}
and $\Delta_p(z,z')$ is the bulk mixed momentum/position space propagator.

Consider as an illustration a portion of the UV brane action
\begin{equation}
S_{UV}[\chi(x,z_{UV})] = - \int d^4 x \sqrt{g_{UV}} \left( \frac{1}{2} 
\lambda_2 \chi^2 
+ \lambda_4 \chi^4 + \ldots \right).
\end{equation}
Decomposing the $\chi$ modes this becomes
\begin{equation}
S_{UV}[\chi_{<} + \chi_{>}]  = - \int d^4 x \sqrt{g} \left(
\frac{1}{2}  \lambda_2 \chi_{<}^2 + 
\frac{1}{2}  \lambda_2 \chi_{>}^2 +
\lambda_4 \left( \chi_{>}^4 + 4 \chi_{<} \chi_{>}^3 +
6 \chi_{<}^2 \chi_{>}^2 + 4 \chi_{<}^3 \chi_{>} + \chi_{>}^4 \right)
+ \ldots \right)
\end{equation}
Now we expand in the terms containing $\chi_>$ and evaluate the path integral
over these modes.  The $\chi_>^2 \chi_<^2$ piece for instance gives a first order
contribution
\begin{eqnarray}
\lefteqn{-i 6 \lambda_4 \int d^4 x \sqrt{g} \chi_{<}(x,z_{UV})^2 
\langle \chi_>(x,z_{UV}) \chi_>(x,z_{UV}) \rangle}  \nonumber \\  &=  & 
-i 6 \lambda_4 \int d^4 x \sqrt{g} \chi_{<}(x,z_{UV})^2 
\int \frac{d^4 p}{(2 \pi)^4} 
(-i) \Delta_p(z_{UV},z_{UV}) \Theta(p) \nonumber \\
&  = &   -i \epsilon \frac{12 \lambda_4}{(4 \pi)^2} \left( \frac{\Lambda}{z_{UV}} \right)^4
\Delta_{i \Lambda /z_{UV}} (z_{UV},z_{UV})
\int d^4 x \sqrt{g} \chi_<(x,z_{UV})^2. 
\end{eqnarray}
This is a  correction to the coefficient $\lambda_2$.
\begin{equation}
\lambda_2 \to \lambda_2 + \epsilon \frac{24 \lambda_4}{(4 \pi)^2} \frac{\Lambda^4}{z_{UV}^4} 
\Delta_{i \Lambda / z_{UV}}(z_{UV},z_{UV}).
\end{equation}

In general, the result of integrating out the modes $\chi_>$ is to generate 
${\cal O}(\epsilon)$ 
corrections to all the coefficients on the UV brane.  We write these correction as
 a local 4D brane action $\tilde{S}_{UV}$.  The result of the integration
is to make the transformation 
\begin{equation}
S_{UV}[\chi(x,z_{UV})] \to S'_{UV}[\chi(x,z_{UV})] = S_{UV}[\chi(x,z_{UV})] +
 \epsilon \tilde{S}_{UV}[\chi(x,z_{UV})].
\end{equation}
Because this is a Wilsonian integration all corrections appear at 1-loop order only.  

Now, we integrate the remaining modes in the $z$ space sliver,
$\chi(x,z_{UV} \le z < z_{UV}(1+\epsilon))$. 
To do this we discretize the space near the effective UV brane and perform the 
path integral over the 4D slice at $z=z_{UV}$.   
We write for the measure
\begin{equation}
\int [d \chi(x,z)] \to \int [d \chi(x,z>z_{UV})] [d\chi(x, z_{UV})] 
\end{equation}
and separate the action as
\begin{equation}
S = S_1[\chi(x,z> z_{UV})] + S_2[\chi(x,z_{UV}),\chi(x,z_{UV}(1+\epsilon ))].
\end{equation}
where $\epsilon z_{UV}$ is the unit of discretization. 
\begin{eqnarray}
S_2[\chi(x,z_{UV}), \chi(x, z_{UV}(1+ \epsilon )] = \frac{\epsilon}{z_{UV}^4} 
\int d^4 x
\Bigg( -\frac{z_{UV}^2}{2} \frac{(\chi(z_{UV}(1+\epsilon )- 
\chi(z_{UV}))^2}{z_{UV}^2 \epsilon^2}
- V(\chi(z_{UV})) \nonumber \\
+ \frac{z_{UV}^2}{2} \eta^{\mu \nu} \partial_{\mu} \chi(z_{UV}) \partial_{\nu} 
\chi(z_{UV})
\Bigg) \nonumber \\
+  \frac{1}{2 z_{UV}^4} \int d^4 x L_{UV}[\chi(z_{UV})] + 
\frac{\epsilon}{2} \tilde{S}_{UV}[\chi(z_{UV})]
\end{eqnarray}

We let
\begin{equation}~\label{transform}
S_{UV}^{\prime \prime}[\chi(x, z_{UV}(1+ \epsilon ) )] =
-i \ln \left( \int [d\chi(x,z_{UV})] e^{i S_2[\chi(x,z_{UV}), 
\chi(x,z_{UV}(1+ \epsilon) )]} \right),
\end{equation}
so that the result of the integration will be an action of the form
\begin{equation}
S = S_1[\chi(x,z> z_{UV})]+ S_{UV}^{\prime \prime}[\chi(x,z_{UV}(1+\epsilon))].
\end{equation}
We will now do the path integral over $\chi(x, z_{UV})$ for infinitesimal $\epsilon$.

We make a change of variables, 
\begin{equation}
\chi(x,z_{UV}) = \chi(x, z_{UV}(1+\epsilon)) - \epsilon \tilde{\chi}(x),
\end{equation}
 and expand in powers of $\epsilon$.   We find
the classical solution $\tilde{\chi}_c(x)$  (in terms of $\chi(x,a+ \epsilon a)$) and 
expand in fluctuations around this value substituting 
$\tilde{\chi} = \tilde{\chi}_c + \eta / \sqrt{\epsilon}$. 
Here
\begin{equation}
\tilde{\chi}_c (x)= -\frac{1}{2} \int d^4 y 
\frac{\delta L_{UV}[\chi(y)]}{\delta \chi(x)} 
\Bigg|_{\chi(x)=\chi(x,z_{UV}+ \epsilon z_{UV})}
+ {\cal O}(\epsilon).
\end{equation}
We expand the quantum fluctuations as $\eta/ \sqrt{\epsilon}$ to obtain a normalized
quadratic term for this field.  This expansion does not create an expansion in inverse powers
of $\epsilon$ as each field $\tilde{\chi}$ is accompanied 
by at least one power of $\epsilon$.
Making this substitution we have
\begin{eqnarray}~\label{transformsolved}
\frac{1}{2} S_{UV}^{\prime \prime} &= & \frac{1}{2} \int d^4 x \frac{1}{z_{UV}^4} 
L_{UV}[\chi(z_{UV}(1+ \epsilon))] \nonumber \\
 &&- \epsilon \int d^4 x \frac{1}{z_{UV}^4} 
\left( V(\chi(z_{UV}(1+\epsilon))) - \frac{z_{UV}^2}{2} \eta^{\mu \nu} 
\partial_{\mu} \chi(z_{UV}(1+ \epsilon) )
\partial_{\nu} \chi(z_{UV}(1+\epsilon)) \right) \nonumber \\
&&+ \frac{\epsilon}{8} \int d^4 x \frac{1}{z_{UV}^4} 
\left( \frac{\delta L_{UV}[\chi(y)]}{\delta \chi(x)}\Bigg|_{\chi=\chi(z_{UV}(1+ \epsilon))} d^4 y \right)^2 
+ \frac{\epsilon}{2} \tilde{S}_{UV}[\chi(z_{UV}(1+\epsilon ))] \nonumber \\
&&+ \frac{1}{i} \ln \Bigg\{ \int  [d \eta] {\rm exp} \Bigg( \frac{i \epsilon}{4}
\int d^4 x  d^4 y d^4 w \eta(y) \eta(w) \frac{1}{z_{UV}^4} \frac{\delta^2 L_{UV}[\chi(x)]}{\delta \chi(y) \delta \chi(w)}
\Bigg|_{\chi=\chi(z_{UV}(1+ \epsilon))}
 \nonumber \\
&& \qquad \qquad  - \frac{i}{2} \int d^4 x \frac{\eta^2}{z_{UV}^4} \Bigg) 
\Bigg\} + {\cal O}(\epsilon^2).
\end{eqnarray}

Now we perform the
quadratic path integral remaining in (\ref{transformsolved}).  
Using  $\ln (\det {\cal O})
=  {\rm tr} \ln {\cal O}$ it evaluates to
\begin{equation}
-\frac{1}{2 i}{\rm tr} \ln \left( 1- \frac{\epsilon}{2} \int d^4 y d^4 w
\frac{\delta^2 L_{UV}[\chi[(x)]}{\delta \chi(y) \delta \chi(w)}
\Bigg|_{\chi=\chi(z_{UV}+ \epsilon z_{UV})} \right) \delta^4 (x-x').
\end{equation}
Keeping only first order in $\epsilon$ and converting the trace to an integral we have
\begin{equation}
 \frac{\epsilon}{4 i VT} \int d^4 x d^4 y d^4 w 
\frac{\delta^2 L_{UV}[\chi(x)]}{\delta \chi(y) \delta \chi(w)}
\Bigg|_{\chi=\chi(z_{UV}+ \epsilon z_{UV})}.
\end{equation}
Where we may take the four volume 
\begin{equation}
\frac{1}{VT} = \int \frac{d^4 p}{(2 \pi)^4} = 
\frac{i \Lambda^4}{ 2 (4 \pi)^2 z_{UV}^4 },
\end{equation}
regulating with the cutoff $\Lambda / z_{UV}$.  

We can now write the effective UV brane as
\begin{eqnarray}~\label{runningequationnot}
\frac{1}{2}  S_{UV}^{\prime \prime} = \frac{1}{2} S_{UV} +
\frac{\epsilon}{8 \sqrt{g_{UV}}} \int d^4 x \left( \frac{\delta S_{UV}}{\delta \chi(x)} 
\right)^2
- \epsilon \int d^4 x \frac{\delta S_{UV}}{\delta g^{\mu \nu}_{UV}(x)} g^{\mu \nu}_{UV}(x) 
\nonumber \\
- \epsilon \int d^4 x \sqrt{g_{UV}} \left( V(\chi) - \frac{1}{2} g^{\mu \nu}_{UV}
\partial_{\mu} \chi \partial_{\nu} \chi \right) 
\nonumber \\
+ \epsilon \frac{\Lambda^4 }{8 (4 \pi)^2} \int d^4y d^4 x 
\frac{\delta^2 S_{UV}}{\delta \chi(y) \delta \chi(x)} + 
\frac{\epsilon}{2} \tilde{S}_{UV}.
\end{eqnarray}
where all fields on the left hand side are evaluated at $\chi(x,z_{UV}(1+\epsilon))$.

Now we rescale the coordinates $x \to x/(1+\epsilon)$, $z \to z/(1+\epsilon)$.
Under this transformation $z_{UV}(1+\epsilon) \to z_{UV}$ and 
$z_{IR} \to z_{IR}/(1+\epsilon)$ but because of the scale symmetry 
{\it there is no additional scaling}.  The effective
action now has the form.
\begin{eqnarray}~\label{action3}
S^{eff}_{bulk}[\chi(x,z)] & =  &\int d^4 x \int_{z_{UV}}^{z_{IR}(1-\epsilon)} dz 
\sqrt{G} \left( \frac{1}{2} G^{MN} 
\partial_M \chi \partial_N \chi - V_{bulk} (\chi) \right), \nonumber \\
S^{eff}_{UV}[\chi(x,z_{UV})] & = & \int d^4 x \sqrt{g_{UV}} L^{eff}_{UV}(\chi(x, z_{UV}), 
\partial_{\mu}, 
g_{UV}^{\mu \nu}), \nonumber \\
S^{eff}_{IR}[\chi(x,z_{IR}(1-\epsilon))] & = & \int d^4 x \sqrt{g'_{IR}} 
L_{IR} (\chi(x, z_{IR}(1-\epsilon)), \partial_{\mu}, 
g_{IR}^{\prime \mu \nu}).
\end{eqnarray}
where $g^{\prime \mu \nu}_{IR}(x) = G^{\mu \nu}(x,z_{IR}(1-\epsilon))$.

We can make this transformation continuous by introducing a mass scale $\mu$ such that 
$\epsilon = \Delta \mu/\mu$.  Taking the limit $\Delta \mu \to 0$
the UV brane obeys
\begin{eqnarray}~\label{runningequation}
\frac{1}{2} \mu \frac{ d S^{eff}_{UV}}{d \mu} =
-\frac{1}{8 \sqrt{g_{UV}}} \int d^4 x 
\left( \frac{\delta S^{eff}_{UV}}{\delta \chi(x)} \right)^2
+ \int d^4 x \frac{\delta S^{eff}_{UV}}{\delta g^{\mu \nu}_{UV}(x)} g^{\mu \nu}_{UV}(x) 
\nonumber \\
+\int d^4 x \sqrt{g_{UV}} \left( V(\chi) - \frac{1}{2} g^{\mu \nu}_{UV}
\partial_{\mu} \chi \partial_{\nu} \chi \right) 
\nonumber \\
- \frac{\Lambda^4 }{8 (4 \pi)^2} \int d^4y d^4 x 
\frac{\delta^2 S^{eff}_{UV}}{\delta \chi(y) \delta \chi(x)} - \frac{1}{2} 
\tilde{S}_{UV}.
\end{eqnarray}
The derivative on the left-hand side acts only on the coefficients of operators 
in $S^{eff}_{UV}$.  We find flow equations for coefficients
by equating coefficients of equivalent operators on both sides of the equality.

The IR brane position flows according
to
\begin{equation}
\mu \frac{d}{d \mu} \tilde{z}_{IR} = \tilde{z}_{IR},
\end{equation}
where we define $\tilde{z}_{IR}(\mu = 1/z_{UV}) = z_{IR}$.  As we take
 $\mu \to 1/z_{IR}$,
$1/\tilde{z}_{IR}$ grows and the proper distance between branes 
$ \sim z_{UV} \ln (z_{UV}/\tilde{z}_{IR} )$ flows to zero.
Note that the running of the radius is a generic 
feature of the RG in compact dimensions but logarithmic scaling is particular to the warped
space picture. 

Equation (\ref{runningequation}) is the central result of this work.  It 
demonstrates that the RG transformations describe 
flows between local theories.  The change in the UV brane action given by 
the right hand side is composed of the action $S_{UV}$ which is taken to be local, a local 
contribution from the bulk action and the piece $\tilde{S}_{UV}$ which 
must be local as it results from integrating out high momentum modes.  
The $\Lambda^4$ term in (\ref{runningequation}) indicates an expected sensitivity to 
the physical cutoff in the flows of UV brane coefficients.  In general, there will also be 
cutoff sensitivity in the term $\tilde{S}_{UV}$.

\section{Fixed Point Behavior}

In this section we discuss the fixed point behavior as obtained from 
equation (\ref{flowequations}).  The flow equations for
UV brane coefficients $C_i$ have the form 
\begin{equation}
\mu \frac{ d C_i}{d \mu} = \beta_i^{cl}(C_j) + \beta^{bulk}_i +
\beta_i^{qu}(C_j, 1/ \tilde{z}_{IR}(\mu) ).
\end{equation}
Here, $\beta^{cl}$ is the classical contribution derived from the first line in
 (\ref{flowequations}) and the quadratic operators in the second line 
 (the mass and kinetic terms).
$\beta^{bulk}$ is the contribution derived from operators of higher than quadratic order 
in the potential in the second line.  This piece
does not depend on $C_i$ but only on bulk coefficients.  
$\beta^{qu}$  is the piece derived from the third line of (\ref{flowequations}).  It contains
all the loop level contributions and any tree level contributions  from $\tilde{S}_{UV}$.
It depends on the IR brane position, $\tilde{z}_{IR}$ but not on bulk coefficients other
than the bulk mass.

First consider the flows of UV brane coefficients 
in the absence of bulk interactions and the corrections in $\beta^{qu}$.
\begin{equation}
\mu \frac{d C_i}{d \mu} = \beta_i^{cl}(C_j).
\end{equation} 
One can find a fixed point, denoted $C_i^{cl*}$, satisfying $
\beta_i^{cl}(C_j^{cl*})=0$.
It corresponds to a noninteracting UV brane configuration.
\begin{equation}
S^{*}_{UV}[\chi(x,z_{UV})]  =  - \int d^4 x \sqrt{g_{UV}}  
\frac{1}{2} \chi(x,z_{UV})  \hat{\lambda}_{2}^{*}(\Box) \chi(x,z_{UV}),
 \end{equation}
 where
\begin{equation}~\label{fixedpoint}
\hat{\lambda}_{2}^{cl*}(\Box) = (4- 2 \nu) + 2 \sqrt{-\Box} 
\frac{J_{\nu-1}(\sqrt{-\Box})}{J_{\nu} (\sqrt{-\Box})},
\qquad \nu = \sqrt{4+m^2},
\end{equation}
and we expand in  
$\Box =  g_{UV}^{\mu \nu} \partial_{\mu} \partial_{\nu}$.

Small deviations from the fixed point obey
\begin{equation}
\mu \frac{d}{d \mu} (C_{i}^{cl}- C_{i}^{cl*}) = 
 \frac{\partial \beta^{cl}_i}{\partial C_j} \Bigg|_{C_j=C_j^{cl*}}
 (C_{j}^{cl}- C_{j}^{cl*}) \equiv \gamma^{cl}_{ij}  (C_{j}^{cl}- C_{j}^{cl*}).
\end{equation}
One can check that $\gamma^{cl}_{ij}$ is an upper triangular matrix with eigenvalues
all positive indicating that $C_i^{cl*}$ is an attractive fixed point.
For a detailed examination of the classical fixed point
behavior we refer the reader to the discussion in \cite{runrs} in which the scale
$\mu$ is identified with the position of the effective UV brane.

Now we add the effect of $\beta^{qu}$ corrections to scaling.  We first
discuss evolution in the RS2~\cite{rs2} limit ($1/z_{IR} \to 0$).  The flow
equations are
\begin{equation}
\mu \frac{d C_i}{d \mu} = \beta_i^{cl}(C_j) + \beta_i^{qu}(C_j, 0)
\end{equation}
Because $\beta_i^{qu}$ is proportional to those coefficients $C_j$ which are zero
at the classical fixed point it is clear
that $\beta_i^{qu} (C_j^{cl*}) = 0$.  The fixed point (\ref{fixedpoint}) is
a fixed point of the quantum theory.  The anomalous dimension matrix
is
\begin{equation}
\gamma_{ij} = \gamma_{ij}^{cl}
+ \frac{\partial \beta^{qu}_i}{\partial C_j} \Bigg|_{C_j=C_j^{cl*}}.
\end{equation}
This matrix is upper triangular implying that the eigenvalues of $\gamma_{ij}$ are equal
to the eigenvalues of $\gamma_{ij}^{cl}$.  Therefore, the fixed point in 
the presence of loops remains attractive.

Allowing for bulk interactions the flow equations become
\begin{equation}
\mu \frac{ d C_i}{d \mu} = \beta_i^{cl}(C_j) + \beta^{bulk}_i +
\beta_i^{qu}(C_j, 0).
\end{equation}
We introduce a formal small parameter $\alpha$ so that 
$\beta^{bulk}_i \sim \alpha$.   Solving for a fixed point 
$C_i^* = C_{i}^{cl*} + \delta C_i^*$ to order $\alpha$ we have
\begin{equation}
\gamma_{ij} \delta C_j^{*} + \beta_i^{bulk} = 0.
\end{equation}
This indicates an ${\cal O}(\alpha)$ shift in the fixed point.  
There will also be a shift of the scaling 
dimensions of ${\cal O}(\alpha)$.  Significantly, for $\alpha$ sufficiently small 
the fixed point will remain attractive in the presence of bulk interactions.

To consider RS1 we turn on $1/z_{IR} \sim {\rm TeV}$.  This 
introduces a relevant parameter in the 
theory appearing in $\beta^{qu}$.
With the scale $\mu$ high above the TeV scale the effect  
on the running is negligible.  We flow towards the fixed point if we are sufficiently close.
At lower scales, $\mu \sim 1/z_{IR}$,  $\tilde{z}_{IR}$ becomes a significant
threshold effect.

At the fixed point, derivatives are dimensionally suppressed not by the cutoff $\Lambda$
but by the curvature $k$.  Our effective theory description will break down at this scale
where the UV brane action becomes nonlocal.  Therefore, we cannot consider the
regime $\Lambda \gg k$ in our RG approach.  It is in the regime $\Lambda > k$ that we
are sensitive to the 5D dynamical structure.  Numerical factors in (\ref{fixedpoint})
suggest that the near fixed point effective theory breaks down at $\Lambda \sim 6 k$ 
allowing us to sample a portion of the extra-dimensional dynamics.

\section{A Loop Diagram Example}

Here, we present a calculation of the one-loop correction to the two point function
 in a theory with a bulk $\chi^4$ interaction.  We compute at the scales 
 $\mu= 1/z_{UV}$, $1/z_{UV} \gg \mu \gg 1/z_{IR}$ and $\mu \sim 1/z_{IR}$.  
 
 \subsection{EFT Calculation with $\mu=1/z_{UV}$}
 
 We  consider the 
 following scalar theory in an RS1 background at the scale $\mu = 1/z_{UV}$.
\begin{eqnarray}
S_{bulk}[\chi(x,z)] & =  &\int d^4 x dz \sqrt{G} \left( \frac{1}{2} G^{MN} 
\partial_M \chi \partial_N \chi - \frac{1}{2} m^2 \chi^2 - g \chi^4 \right), 
\nonumber \\
S_{UV}[\chi(x,z_{UV})] & = & - \int d^4 x \sqrt{g_h}  \frac{1}{2} 
\lambda_{2}(\mu \sim 1/z_{UV})
\chi^2, \nonumber \\
S_{IR}[\chi(x,z_{IR})] & = & 0.
\end{eqnarray}
where we take $m,g < 1$ and $\lambda_{2}(\mu \sim 1/z_{UV}) \sim 1$.

The propagator with 4-momentum $q$ is 
\begin{equation}~\label{propagator}
\Delta_q(z,z') =  z^2 z^{\prime 2} \left( \Delta_{>}(z,z') \theta(z - z') + 
\Delta_{<}(z,z') \theta(z' - z) \right),
\end{equation}
where $\Delta_{>}(z,z') = \Delta_{<}(z',z)$ and
%\end{equation}
\begin{eqnarray}
\Delta_{>}(z,z') & = & C A(qz',qz_{UV}) B(qz,qz_{IR}),  \\
A(qz',qz_{UV}) & = & (\tilde{N_{\nu}}(q z_{UV})- \frac{\lambda_{2}}{2}
N_{\nu}(q z_{UV}) ) J_{\nu}(qz')-   (\tilde{J_{\nu}}(q z_{UV})- \frac{\lambda_{2}}{2}
J_{\nu}(q z_{UV}) ) N_{\nu}(qz'), \nonumber \\
B(qz,qz_{IR}) & = & \tilde{N_{\nu}}(q z_{IR}) J_{\nu}(q z) - 
\tilde{J_{\nu}}(q z_{IR}) N_{\nu} (q z),  \nonumber \\
C^{-1} & = &\frac{2}{\pi} (\tilde{N_{\nu}}(q z_{UV})- \frac{\lambda_{2}}{2}
N_{\nu}(q z_{UV}) ) \tilde{J}_{\nu}(qz_{IR})-   (\tilde{J_{\nu}}(q z_{UV})- \frac{\lambda_{2}}{2}
J_{\nu}(q z_{UV}) ) \tilde{N}_{\nu}(qz_{IR}), \nonumber
\end{eqnarray}
with
\begin{equation}
\tilde{J_{\nu}}(qz) = (1-\frac{\nu}{2} ) J_{\nu}(qz) + \frac{q z}{2}  J_{\nu-1}(qz).
\end{equation}
and similarly for $\tilde{N_{\nu}}$.  A calculation of this scalar propagator
is similar to that found in~\cite{grinstein,giddings}.  

The one loop contribution to the two-point function with external points evaluated
at $z=z_{IR}$ is
\begin{equation}
\langle \chi(q, z_{IR}) \chi(-q, z_{IR}) \rangle = -i \Delta_q(z_{IR},z_{IR}) +
12 g \int_{z_{IR}}^{z_{UV}} dz \frac{1}{z^5} \Delta_q(z_{IR},z)^2 \int 
\frac{d^4p}{(2 \pi)^4} \Delta_p(z,z).
\end{equation}
We will set external momentum $q=0$ and
 cutoff the momentum integral at the scale $\Lambda / z$.  It is natural to take
$\Lambda \sim {\cal O}(M_{Pl})$.  However, because our purpose is only to illustrate
the effective theory we will greatly simplify 
the calculation by taking the limit $\Lambda \ll 1/z_{IR}$.  
In this limit the propagator has the simple momentum independant form.
\begin{equation}
\Delta_{>}(z,z') \approx \frac{1}{2 \nu} \left(\frac{z'}{z} \right)^{\nu} 
-\frac{1}{2 \nu} \frac{2-\nu}{2+\nu} \left( \frac{z' z}{z_{IR}^2} \right)^{\nu}.
\end{equation}
With this simplification we can perform an analytical 
evaluation of the graph to obtain
\begin{equation}~\label{fulltheory}
\langle  \chi(q, z_{IR}) \chi(-q, z_{IR}) \rangle \approx 
\frac{-i z_{IR}^4}{2+\nu} +
\frac{ i 3 g z_{IR}^4 \Lambda^4}{2(4 \pi)^2 (2+\nu)^3 \nu^2} \left( 1+\frac{3 \nu}{2}
\right).
\end{equation}
where we neglect positive powers of the negligible quantity $z_{UV}/z_{IR}$.
Though this result is highly dependent on the regularization scheme and very sensitive to
ultraviolet physics we will use a consistent regulator in the low scale effective
theory calculations to follow.

\subsection{EFT Calculation with $1/z_{UV} \gg \mu \gg 1/z_{IR}$}

Next, we compute the ${\cal O}(g)$ correction to the two point function 
at the scale $\mu$ such that $1/z_{UV} \gg \mu  \gg 1/z_{IR}$.
At this scale the action has the form
\begin{eqnarray}
S_{bulk}^{eff}[\chi(x,z)] & =  &\int d^4 x \int_{z_{UV}}^{\tilde{z}_{IR}(\mu)} dz \sqrt{G} 
\left( \frac{1}{2} G^{MN} 
\partial_M \chi \partial_N \chi - \frac{1}{2} m^2 \chi^2 - g \chi^4 \right), 
\nonumber \\
S_{UV}^{eff}[\chi(x,z_{UV})] & = & -  \int d^4 x \sqrt{g_{UV}} \left( 
\frac{1}{2} \lambda_2(\mu) \chi^2 
+ \lambda_4(\mu) \chi^4 + \ldots \right), \nonumber \\
S_{IR}^{eff}[\chi(x,\tilde{z}_{IR}(\mu))] & = & 0.
\end{eqnarray}
Here $\ldots$ indicates the presence of operators with higher derivatives and higher
powers of the field $\chi$.  

We need to evolve 
the quadratic and quartic UV brane coefficients from the scale $\mu_{0}=1/z_{UV}$ to $\mu$ 
according to the flow equations given in (\ref{runningequation}).  These are
\begin{eqnarray}~\label{flowequations}
\mu \frac{d}{d\mu} \lambda_2 & = & \frac{1}{2} \lambda_2^{2} 
- 4 \lambda_2 - 2 m^2 - 
\frac{6 \Lambda^4}{(4 \pi)^2} \lambda_4 - 
\frac{24 \lambda_4 \Lambda^4}{(4 \pi)^2} 
\left( \frac{\Delta_{i \Lambda / z_{UV}} (z_{UV},z_{UV}) }{z_{UV}^4} \right) \\
\mu  \frac{d}{d\mu} \lambda_4 & = & - 4 \lambda_4 + 2 \lambda_2 
\lambda_4 - 2 g, ~\label{lambda4}
\end{eqnarray}
where we have not included corrections that will be unimportant
to first order in $g$.   

In the limit $\Lambda \ll 1/z_{IR}$ the propagator appearing in (\ref{flowequations})
simplifies.  The flow equation becomes
\begin{equation}
\mu \frac{d}{d\mu} \lambda_2  =   \frac{1}{2} \lambda_2^{2} 
- 4 \lambda_2 - 2 m^2 - 
\frac{6 \Lambda^4}{(4 \pi)^2} \lambda_4 - 
\frac{24  \Lambda^4}{(4 \pi)^2 2 \nu} \lambda_4 +
 {\cal O} \left( \left( \frac{1}{\mu z_{IR}} \right)^{2 \nu} \right).
\end{equation}
In our $\mu \gg 1/z_{IR}$ limit we may ignore the small $(1/ \mu z_{IR})^{2 \nu})$
correction.  

We can find the fixed point of these flows.  To order $g$ it is
\begin{eqnarray}~\label{lambda2star}
\lambda_2^{*} &  = & 2(2+ \nu) + \delta \lambda_2^{*} = 2(2+ \nu) + 
\frac{3 \Lambda^4 g (\nu +2)}{2(4 \pi)^2 \nu^2 (\nu+1)}\\
\lambda_4^{*} & =  & \frac{g}{2+2 \nu}.
\end{eqnarray}
This is an attractive fixed point.  Given our initial UV brane configuration at 
$\mu=1/z_{UV}$ we will flow to this fixed 
point if $g$ is sufficiently small and $\lambda_2(\mu=1/z_{UV})$ is sufficiently close
to $2(2+\nu)$. We will assume that we are in this attractive regime.   Deviations from
these fixed point values at $1/z_{UV} \gg \mu$
will be suppressed by positive powers of $\mu z_{UV} \sim {\cal O}(\mu/ M_{Pl})$.  
We will ignore these deviations in the evaluation of the 2-point function.

We can now calculate the ${\cal O}(g)$ 2-point function in this effective theory.  For
the propagator we use that of equation (\ref{propagator}) but with $\lambda_2 \to 2(2+\nu)$
and $z_{IR} \to z_{IR} \mu$.
We will treat ${\cal O}(g)$ contributions to the quadratic brane coefficients perturbatively.  
The two point function is now
\begin{eqnarray}
\frac{1}{\mu^4} \langle \chi(\frac{q}{\mu}, z_{IR}\mu) 
\chi(-\frac{q}{\mu},z_{IR} \mu) \rangle 
& = & -i \frac{1}{\mu^4}\Delta_{q/\mu}(z_{IR} \mu, z_{IR} \mu) \nonumber \\
&& +
12 \frac{g}{\mu^4} \int_{z_{IR}\mu}^{z_{UV}} dz \frac{1}{z^5} \Delta_{q/\mu} (z_{IR}\mu,z)^2 
\int \frac{d^4 p}{(2 \pi)^4} \Delta_p (z,z) \nonumber \\
& & + \frac{1}{\mu^4 z_{UV}^4} \Bigg(\frac{i \delta \lambda_2^{*}}{2} 
\Delta_{q/\mu}(z_{IR}\mu,z_{UV})^2
\nonumber \\ &&+ 6 \lambda_4^{*} \Delta_{q/\mu}(z_{IR}\mu,z_{UV})^2 
\int \frac{d^4 p}{(2 \pi)^4} \Delta_p(z_{UV},z_{UV}) \Bigg).
\end{eqnarray}
The overall factor of $\mu^{-4}$ is required to match the 2-point correlator in 
4-momentum space.

We solve for zero external momentum.  The leading contribution from the UV brane
interactions cancels with a piece from the bulk loop integral and we find
\begin{equation}~\label{halftheory}
\frac{1}{\mu^4} \langle  \chi\frac{q}{\mu}, z_{IR}\mu) 
\chi(-\frac{q}{\mu}, z_{IR} \mu) \rangle 
\approx \frac{-i z_{IR}^4}{2+\nu} +
\frac{ i 3 g z_{IR}^4 \Lambda^4}{2(4 \pi)^2 (2+\nu)^3 \nu^2} 
\left( 1+\frac{3 \nu}{2} \right) .
\end{equation}
This matches the result of equation (\ref{fulltheory}).

\subsection{EFT Calculation with $\mu \sim 1/z_{IR}$}

Now we compute the first order correction to the two point function in 
the low energy effective theory at $\mu \sim 1/z_{IR}$.  At this scale
the IR brane position $\tilde{z}_{IR}(\mu = 1/z_{IR}) = z_{UV}$ so that
the bulk contribution to the action is negligible.
The
effective theory is then approximately four dimensional.  It has the form
\begin{equation}
S^{eff} \approx S_{UV}^{eff}[\chi(x,z_{UV})] = 
- \int d^4 x \sqrt{g_{UV}} \left( \frac{1}{2} \lambda_2(\mu \sim 1/z_{IR}) \chi^2 
+ \lambda_4(\mu \sim 1/z_{IR}) \chi^4 + \ldots \right).
\end{equation}

We must evolve $\lambda_2$ and $\lambda_4$ 
from $\mu_{0}=1/z_{UV}$ to $\mu = 1/z_{IR}$ with the flow equations 
(\ref{flowequations}) and
(\ref{lambda4}).  In the last subsection 
we ignored corrections of order $(1/\mu z_{IR})$.  Here we must include them. Writing 
$\lambda_2 = \lambda_2^{*} + \delta \lambda_2$, with $\lambda_2^{*}$ given
by (\ref{lambda2star}) the flow equation
for $\delta \lambda_2$ is
\begin{equation}
\mu \frac{d}{d \mu} \delta \lambda_2 = 2 \nu \delta \lambda_2 +
\frac{24 \lambda_4^{*} \Lambda^4}{(4 \pi)^2 2 \nu} \frac{2-\nu}{2+\nu}
\left( \frac{1}{\mu z_{IR}} \right)^{2 \nu}.
\end{equation}
At the scale $\mu\sim 1/z_{IR}$ find
\begin{equation}
\delta \lambda_2(\mu \sim 1/ z_{IR}) = 
- \frac{\Lambda^4 \lambda_4^{*} (2-\nu)}{2(4 \pi)^2  \nu^2 (2+ \nu)} + 
{\cal O}((\mu z_{UV})^{2 \nu}).
\end{equation}
 We will ignore the ${\cal O}(\mu z_{UV})$ deviations.   
 The effective theory coefficients at $\mu \sim 1/z_{IR}$ are then
 \begin{eqnarray}
 \lambda_2(\mu \sim 1/z_{IR}) & = & 2(2+\nu)+ \delta \lambda_2^* + 
 \delta \lambda_2(\mu) = 2(2+\nu) +\frac{3 \Lambda^4 g
  (\nu^2+5 \nu +2)}{2 (4 \pi)^2 \nu^2 (2+\nu) (1+\nu)} \\
 \lambda_4(\mu \sim 1/z_{IR}) & = & \lambda_4^* = \frac{g}{2+2\nu}.
\end{eqnarray}

We now calculate the two point function to ${\cal O}(g)$ in this effective theory.
It is
\begin{eqnarray}
\frac{1}{\mu^4} \langle \chi(\frac{q}{\mu},z_{IR} \mu) \chi(-\frac{q}{\mu}, z_{IR}\mu ) \rangle 
&= &
 -i \frac{1}{\mu^4} \tilde{\Delta}_{q/\mu}(z_{IR}\mu,z_{IR}\mu)  \nonumber \\
&& + 
\frac{1}{\mu^4 z_{UV}^4 } \Bigg( \frac{i (\delta \lambda_2^*+ \delta \lambda_2(\mu))}{2} 
\tilde{\Delta}_{q/\mu}(z_{IR}\mu,z_{UV})^2 \nonumber \\
&& + 6 \lambda_4(\mu) \tilde{\Delta}_q(z_{IR}\mu ,z_{UV})^2 \int \frac{d^4p}{(2 \pi)^4} 
\tilde{\Delta}_p (z_{UV},z_{UV})
\Bigg).
\end{eqnarray}
where $\tilde{\Delta}_q$ is the ${\cal O}(g^0)$ propagator.  The momentum integral
is cutoff at the scale $\Lambda/z_{UV}$.

In our approximation $\Lambda \ll 1/z_{IR}$ the  propagator 
is simply $\tilde{\Delta} \approx z_{IR}^4 /(2+ \nu)$.  Evaluating the momentum
integral the expression becomes
\begin{equation}
\frac{1}{\mu^4} \langle \chi(\frac{q}{\mu}, z_{IR} \mu) \chi(-\frac{q}{\mu}, z_{IR} \mu) \rangle \approx 
\frac{-i z_{IR}^4 }{2+\nu}+ \frac{i 3 g z_{IR}^4 \Lambda^4}{2(4 \pi)^2 (2+\nu)^3 \nu^2}
\left( 1+ \frac{3 \nu}{2} \right).
\end{equation}
Again we see that this matches with equation (\ref{fulltheory}).

We stress that though we have worked in the limit $\Lambda \ll 1/z_{IR}$ to simplify the 
comparison to the full theory result
it was not conceptually necessary to make this approximation. 
 In fact, there is no major obstacle to working
in the region $\Lambda \sim 1/z_{UV} = k$ in the low energy EFT with 
$\mu \sim 1/z_{IR}$.

\section{Discussion}

We have derived the $\beta$ functions for couplings of operators of a scalar
field in the compactified RS background.  We see that bulk and IR brane couplings do 
not suffer RG transformations while UV brane couplings flow according to 
\begin{equation}
\mu \frac{d C_i}{d \mu} = \beta_i  ( C_j, 1/ (\mu z_{IR}) )
\end{equation}
where the $\beta_i$ can be read from equation 
 (\ref{runningequation}).  The IR brane position transforms as
 \begin{equation}
 \mu \frac{d \tilde{z}_{IR}}{d \mu} = \tilde{z}_{IR}.
 \end{equation}
scaling the radius of the extra dimension to zero in the IR.
There is an attractive fixed point in this 
theory corresponding to the RS2 limit, $z_{IR} \to \infty$.
 
 The RG evolution that we have found stands alone on the gravity side and does not make
 use of a duality relationship.  However, properties of the dual theory are evident.
 There is a fixed point and a relevant mass parameter identifying the IR brane with a breaking 
 of scale symmetry.  The scaling behavior of UV brane couplings away from the fixed point
can be identified with the scaling of coefficients of operators in the CFT dual that 
explicitly break the conformal symmetry.  The quantum corrections to the
flow equations we have found correspond to corrections to the large $N$ 
CFT side flows.  

Next, we note that it is a useful calculational tool to work in the low energy effective 
field theory.  
In any bulk RS loop diagram we must perform integrals over internal momentum
and integrals over extra-dimensional coordinates.   This involves challenging integration
of Bessel's functions over a wide range of scales.  Momentum flowing along internal 
propagator lines connecting two bulk vertices is cut off at a range of scales that
is difficult to implement in cutoff regularization.   In our EFT approach, high momentum
modes are integrated out in slices of $z$ space where it is clear how to implement the cutoff.  

Power counting
is straightforward in the low energy EFT.   At sufficiently low scales the RS effective theory
can be described 
as an approximately flat effectively TeV sized extra dimensional theory or (at lower scales) a 
four dimensional theory.  Here, operators of higher dimension
are suppressed naturally by the warped down cutoff scale.  Diagrams 
are simple to calculate in these scaled down theories where the propagator has a simple
flat space form.  To work in the low energy theory we must compute the effective 
UV brane couplings at low scales.  However, near the attractive fixed point, RG 
flows for these couplings are simple.
In this paper we have performed the effective theory calculation
of the 1-loop correction to the two point function in a scalar theory with a bulk quartic
interaction at three RG scales
showing that they match. 

Finally, it will be interesting and relevant to develop along the lines presented here, the RG 
formalism appropriate for higher spin fields propagating in the RS background.  We expect
the RG will have similar features to the scalar case:  a scaling only of UV brane coefficients 
and a running of the radius.  This program was begun at tree level for integer spin fields 
in \cite{higherspin}.   Extending this approach to the loop level will allow for
its application to gauge coupling unification.  
An understanding of the RG in the presence of dynamical gravity would allow for a complete
EFT loop level examination of the Goldberger-Wise stabilization
\cite{goldwise} of the RS1 hierarchy.

\Acknowledgements

I would like to thank Thomas Becker, Patrick Fox, Elie Gorbatov, Emanuel Katz, 
Ben Lillie, Michael Peskin and Raman Sundrum for helpful conversations.

\end{document}